\documentclass[nofootinbib,superscriptaddress]{revtex4-1}

\usepackage{amsmath}
\usepackage{amsfonts}
\usepackage{graphicx}

\begin{document}

\title{UV and IR quantum-spacetime effects for the Chandrasekhar model}

\author{Giovanni AMELINO-CAMELIA}
\affiliation{Dipartimento di Fisica, Universit\`a di Roma ``La Sapienza"\\
and Sez.~Roma1 INFN, P.le A. Moro 2, 00185 Roma, Italy}

\author{Niccol\`o LORET}
\affiliation{Dipartimento di Fisica, Universit\`a di Roma ``La Sapienza"\\
and Sez.~Roma1 INFN, P.le A. Moro 2, 00185 Roma, Italy}

\author{Gianluca MANDANICI}
\affiliation{Universit\`a degli Studi di Bergamo, Facolt\`a di Ingegneria,\\
Viale Marconi 5, 24044 Dalmine (Bergamo) Italy.}

\author{Flavio MERCATI}
\affiliation{Departamento de F\`isica T\`eorica, Universidad de Zaragoza,\\ Zaragoza 50009, Spain}

\begin{abstract}
We modify the Chandrasekhar model of white dwarfs by
introducing novel momentum-space features that  characterize the analysis
of some quantum-spacetime scenarios. We find that the rather standard ultraviolet effects
of spacetime quantization can only be significant in a regime where the Chandrasekhar model
anyway lacks any contact with observations. But a new class of
quantum-spacetime effects inspired by
the mechanism of ``ultraviolet/infrared mixing" could be relevant for white dwarfs
whose mass is roughly half the mass of the Sun, some of which are described in the
literature as ``strange white dwarfs". We also offer a preliminary argument suggesting
that Planck-scale (ultraviolet) effects could be significant in cases where
ultra-high densities are present, even when the relevant star is still gigantic
in Planck-length units.
\end{abstract}

\maketitle

\section{Introduction}
Over the last decade a few experimental/observational contexts have been
identified in which remarkably some ultrasmall effects
introduced at the Planck scale,
of the type that can be of interest for
the study of the quantum-gravity problem, could be tested
(see, {\it e.g.},
Refs.~\cite{grbgac,kifune,biller,gacNature1999,gacPiranPRD,jaconature,PiranNeutriNat,gacPolonpapLRR}).
This had a significant impact on the perspective adopted by many of those
studying the quantum-gravity problem, especially in light of the fact that
up to the mid 1990s it was instead assumed as a given that
the smallness of the characteristic distance scale of quantum gravity,
the ``Planck length" $L_P \sim 10^{-35}m$, would render
these effects forever inaccessible.
While this is certainly exciting, it should be stressed that
the very few examples of  tests of Planck-scale physics that have been identified so far
provide us only with opportunities to explore the Planck-scale regime in a rather
limited way.
In particular, these tests are predominantly focused on contexts that allow one to
neglect the "gravity aspects" of the quantum-gravity problem: one tests candidate quantum properties
of spacetime, linked to gravity only by its role in motivating the quantum-spacetime hypothesis,
in situations where the geometrodynamics of spacetime can indeed be ignored.
We have already argued elsewhere~\cite{grf2010honor} that it might be proper
to at least explore some aspects of the interplay between quantum-spacetime effects and gravity.
We here take as starting point an analysis, which fits perfectly the strategy
we are advocating, reported by Camacho
 in Ref.~\cite{Camacho}. This concerns how the
 Chandrasekhar model~\cite{Chandrasekhar} of white dwarfs could be modified introducing a class of
 energy-momentum (on-shell/dispersion) relation
 that have been of interest in the quantum-spacetime literature.

The results of the preliminary analysis reported by Camacho in  Ref.~\cite{Camacho}
are particularly intriguing since they
provide evidence of a striking amplification of Planck-scale effects
in the limit in which the mass of the white
dwarf approaches the Chandrasekhar limit.
However, due to the preliminary nature of Camacho's investigation, it remains unclear whether this
amplification could be turned into a resource from a phenomenology perspective,
and it was not clear which structures within the  Chandrasekhar model
were responsible for that surprising result.
A first follow-up study was reported in Ref.~\cite{sethNANE}, analyzing in some detail a few
more aspects of the relevant problem, showed  that the amplification of Planck-scale effects
reported by Camacho is unlikely to provide the basis for a viable phenomenological programme,
at least not in the context of the Chandrasekhar model and of its Newtonian description
of gravitational effects.

We here extend rather significantly the scope of these investigations
of candidate quantum-spacetime effects for the Chandrasekhar model.
One of the issues we consider is the one concerning the
possibility that spacetime symmetries might be deformed, in the sense
of the ``doubly-special relativity"
framework~\cite{gacdsr1a,gacdsr1b,dsrO1,dsrO2,dsrnature,dsrO3,dsrO4,rainbowDSR,gacdsrrev2010}:
both Ref.~\cite{Camacho} and Ref.~\cite{sethNANE} introduced Planck-scale modifications of
the dispersion relation
in such a way that the underlying spacetime symmetries would be broken, with the emergence of a preferred frame, whereas here we included the possibility of a modified measure of integration over momentum-space
variables, as required by attempts to deform rather than break Lorentz invariance.

Even within the more general framework we consider we still find that
the type of ultraviolet/Planck-scale effects considered in Refs.~\cite{Camacho,sethNANE}
cannot produce observably large effects
in the context of the Chandrasekhar model and of its Newtonian description
of gravitational effects. But we notice that the results nonetheless provide a valuable
intuition by showing that, whereas naively one might expect quantum-spacetime effects
to only matter in the idealized case of bodies of Planck-length size,
in the context of the Chandrasekhar model one finds that quantum-spacetime effects
would be important in a regime where the body has
extremely high density but not necessarily Planck-length size.
This does not open any windows of phenomenological relevance for studies of white dwarfs,
but may motivate a dedicated search of contexts where instead the same feature
is phenomenologically relevant. Since we feel this issue is of substantial interest
we also preliminarily investigate (in Section~\ref{tolman}) its interplay with
 general-relativistic
effects.

We also study the implications
for the Chandrasekhar model of a class of candidate quantum-spacetime effects
which was not considered in Refs.~\cite{Camacho,sethNANE}.
These are effects introduced in some scenarios
with ``ultraviolet/infrared mixing",
where the quantum structure of spacetime, besides being relevant at very short distance scales,
is also relevant in some far-infrared regime. For this case we do find a window
of opportunity for an associated phenomenology, which in particular might eventually lead
  to a novel interpretation
of the so-called ``strange white dwarfs".

\section{Chandrasekhar model with UV modifications of the dispersion relation}\label{sectiontwo}
It is convenient for us to start by summarizing briefly the argument
proposed by Camacho in Ref.~\cite{Camacho}, so that the perspective from which
we are planing to contribute will be clearer.
The analysis of Ref.~\cite{Camacho}
adopts a class of candidate modifications of the dispersion relation,
parametrized\footnote{Camacho considered the more general case
of a quantum-gravity correction of the form $\eta \vec{p}^2 ({E}/{E_p})^\alpha$
with the additional parameter $\alpha$ giving the power of the leading-order
quantum-gravity correction. For simplicity we here focus on the most studied scenario,
in which $\alpha =1$. Since we are investigating whether there are any chances of
significant effects it makes sense to focus on a definite and well-studied case.}
 by a parameter $\eta$  in terms
of the Planck scale $E_p$ (we use units such that $c=\hbar=1$)
\begin{equation}
E^2-m^2 = \vec{p}^2 + \eta \vec{p}^2 \left(\frac{E}{E_p}\right) \label{MDR},
\end{equation}
which has been much studied in the literature (see, {\it e.g.},
Refs.~\cite{aemn1,plb1997,grbgac,gambinipullin,urrutiaPRL,gacdsr1a,gacdsr1b}).
And Ref.~\cite{Camacho}
observes that such a modification of the dispersion relation would primarily affect
the analysis of the Chandrasekhar model for white dwarfs
by producing a correspondingly modified expression
for the zero point energy of the system:
\begin{eqnarray}
E_0=\frac{2V}{8\pi^3}\int_0^{p_F}  dp ~ 4\pi p^2\left[ \sqrt{p^2 + m_e^2}
\left(1+\frac{\eta}{2}\frac{p^2}{p^2
+m_e^2}\left(\frac{\sqrt{p^2+ m_e^2}}{E_p}\right)\right)-m_e\right] \label{Ezerocorr1} ~,
\end{eqnarray}
 where $m_e$ is the electron mass and $p_F = (3\pi^2 N/V)^{\frac{1}{3}}$ is the Fermi momentum.

 From Eq.~(\ref{Ezerocorr1}) one finds
\begin{eqnarray}
E_0=\frac{m_e^4 V}{\pi^2}\left(f(x_F)+\frac{\eta}{2}\left(\frac{m_e
}{E_p}\right)g(x_F)\right)\label{Ezerocorr}
\end{eqnarray}
in terms of the dimensionless variable $x_F \equiv \frac{p_F}{m_e}$ and with
\begin{eqnarray}
f(x_F)&=&\frac{x_F}{4}(x_F^2+1)^{3/2}-\frac{x_F}{8}\sqrt{x_F^2+1}
-\frac{1}{8}\ln\left(x_F+\sqrt{x_F^2+1}\right)-\frac{1}{3}x_F^3 ~,\label{effe}\\
g(x_F)&=& \int_0^{x_F}x^4 dx = \frac{1}{5} x_F^5 ~.
\end{eqnarray}

The result (\ref{Ezerocorr}) was then used by Camacho~\cite{Camacho}
to study the implications of the quantum-gravity effects
for the pressure due to Pauli repulsion
in the ultrarelativistic $(x_F>>1)$ limit:
\begin{equation}
P_0 = -\frac{\partial E_0}{\partial V} \simeq \left(\frac{m_e^4}{12\pi^2}\right)\left(x_F^4-x_F^2
+ \frac{4}{5}\eta\frac{m_e}{E_p}x_F^5\right) ~.\label{pizerorelcorr}
\end{equation}

This is rather significant, at least conceptually, since within
 the Chandrasekhar model~\cite{Chandrasekhar}
 the stability of white dwarfs is due to the equilibrium between
gravitational pressure and Pauli repulsion.
Therefore the result of Eq.~(\ref{pizerorelcorr})
should balance the gravitational pressure, which in the Chandarsekhar model is
described using Newtonian gravity:
\begin{equation}
P_0=\frac{G}{4\pi}\left(\frac{8 m_P}{9\pi}\right)^2 m_e^4\frac{\bar{M}^2}{\bar{R}^4}\label{Pgrav},
\end{equation}
where  $m_P$ is the proton mass and
\begin{equation}
\bar{M}=\frac{9\pi }{8 m_P}M ~, \qquad \bar{R}= m_e R  ~.
\end{equation}
Since $ x_F = {\bar M}^{\frac{1}{3}}/\bar{R} $, Camacho's quantum-gravity-modified pressure-balance equation
therefore takes the form
\begin{eqnarray}
K\frac{\bar{M}^{4/3}}{\bar{R}^4}-K\frac{\bar{M}^{2/3}}{\bar{R}^2}
+\frac{4}{5}\eta\frac{m_e}{E_p}K\frac{\bar{M}^{5/3}}{\bar{R}^5}
=K'\frac{\bar{M}^2}{\bar{R}^4}~, \label{equaz}
\end{eqnarray}
in which $K=\left(\frac{m_e^4}{12\pi^2}\right)$
and $K'=\frac{G}{4\pi}\left(\frac{8m_P}{9\pi}\right)^2 m_e^4$.

Solving Eq.~(\ref{equaz}) for the radius one finds
\begin{equation}
\bar{R}=\bar{M}^{\frac{1}{3}}\sqrt{1-\frac{K'}{K}\bar{M}^{\frac{2}{3}}}
+\frac{4}{5}\eta\frac{m_e}{E_p}\frac{\bar{M}^{\frac{1}{3}}}{2(1
-\frac{K'}{K}\bar{M}^{\frac{2}{3}})} ~,\label{dodici}
\end{equation}
which is most intelligibly characterized in terms of a correction to
the Chandrasekhar result:
\begin{equation}
R = R_{Chan}\left(1+\frac{2}{5}\eta\frac{m_e}{E_p}\frac{1}{(1
-(\frac{M}{M_0})^{\frac{2}{3}})^{\frac{3}{2}}}\right) ~,\label{Camachos}
\end{equation}
where
\begin{equation}
R_{Chan}=\frac{3}{2}\frac{\pi^{\frac{1}{3}}}{m_e}\left(\frac{M}{m_p}\right)^{\frac{1}{3}}\sqrt{1-\left(\frac{M}{M_0}\right)^{\frac{2}{3}}}
\end{equation}
is the Chandrasekhar radius and
$$M_0=(\frac{9\pi}{8 m_p})^2(\frac{1}{3\pi G})^{\frac{3}{2}}$$
is the so-called Chandrasekhar limit, giving the maximum stable mass for a white dwarf in the
Chandrasekhar model.

The remarkable aspect of Camacho's result (\ref{Camachos}) is the huge amplification of the
quantum-gravity effect for $M \rightarrow M_0$: the quantum-gravity correction has the inevitable
factor with Planck-scale suppression $\frac{m_e}{E_p}$ (and this is of order $\sim 10^{-22}$)
but in this instance it also involves
a factor $(1-(\frac{M}{M_0})^{\frac{2}{3}})^{-\frac{3}{2}}$ which formally diverges
as  $M \rightarrow M_0$.
For some value of $M$ that is very close but finitely smaller than $M_0$ the correction
is of $O(1)$.
This result is evidently very significant from a quantum-gravity perspective, and we shall provide
further elements to underline this significance here.

From the conceptual perspective it is however interesting to investigate whether it is
possible that
this unexpected feature is a peculiarity of scenarios with broken Lorentz symmetry,
or could be also a feature of scenarios with deformation of Lorentz symmetry in the sense
of ``doubly-special relativity"~\cite{gacdsr1a,gacdsr1b}. Both in Camacho's original paper~\cite{Camacho}
and in a follow up study~\cite{sethNANE} there was no evidence that this feature might
be prevented in symmetry-deformation scenarios, but we shall here show that this is a possibility.

Most importantly, one would hope that this unexpected and striking feature could be exploited
for experimental searches of quantum-gravity effects, in the spirit of the recent literature
on ``Quantum-Gravity Phenomenology"~\cite{grbgac,kifune,biller,gacNature1999,gacPiranPRD,jaconature,PiranNeutriNat,gacPolonpapLRR}).
From this perspective Camacho~\cite{Camacho} focused his attention preliminarily on
some white dwarfs with mass of roughly half a solar mass.
One of our objectives is a more in-depth
analysis of the possibility of using data on such white dwarfs
 as opportunities to place meaningful constraints on the model.
And our first observation on this point
is that these white dwarfs cannot be described in the ultrarelativistic
regime, on which Camacho focused, since for them the condition $x_F>>1$ is not satisfied.
This will lead us to generalize Camacho's analysis to include proper handling of the nonrelativistic
regime.
And this in turn will lead us to raise the possibility (not considered in Refs.~\cite{Camacho,sethNANE})
of considering modifications of the dispersion relation which are also motivated
by the quantum-spacetime/quantum-gravity, but for scenarios with ``ultraviolet/infrared mixing"
so that there is a better chance of finding tangible consequences for white dwarfs
in the not-ultrarelativistic regime.

We shall also consider (however preliminarily, in Section~VI) some possible implications
found from including some
general-relativistic corrections.
The only study that so far has extended Camacho's original observation
was reported in Ref.~\cite{sethNANE} and focused mainly on
improving the picture advocated by Camacho in the direction of obtaining
a more
robustly quantitative analysis, including in particular the effects of
a non-homogeneous mass-density
distribution within the star,
but still focusing on a Newtonian description of gravitational effects.

\subsection{Deformed integration measure}
Both Ref.~\cite{Camacho} and Ref.~\cite{sethNANE}
considered modifications of the Chandrasekhar model implied by quantum-gravity scenarios
with Planck-scale modifications of the dispersion relation. But both papers did not
investigate the possible differences between scenarios with breakdown of Lorentz
symmetry and scenarios with deformed Lorentz symmetry.

The proposal~\cite{gacdsr1a,gacdsr1b} of deformed Lorentz symmetry, in the sense of DSR (``doubly-special
relativity"), has generated rather significant interest and a relatively large literature
(see, {\it e.g.},
Refs.~\cite{gacdsr1a,gacdsr1b,dsrO1,dsrO2,dsrnature,dsrO3,dsrO4,rainbowDSR,gacdsrrev2010}).
Crucial for our purposes is the fact,
already established with the first results~\cite{gacdsr1a,gacdsr1b} of this research programme,
that modified dispersion relation of the type (\ref{MDR}) can be introduced
in a deformed-symmetry scenario, {\underline{but only if accompanied by a modification
of}} {\underline{the law of composition of momentum}}.
The studies reported in
Ref.~\cite{Camacho} and Ref.~\cite{sethNANE} focus on the fact that there is a clear
explicit role for the dispersion relation in the Chandrasekhar analysis,
while there is no explicit role for the law of composition of momentum.
However, we want to here stress that, as shown already in Refs.~\cite{gacmajid,kowaMEASURE},
the law of composition of momenta affects the rules of integration over energy-momentum
space, and these are crucially relevant for the Chandrasekhar analysis.

The considerations reported in Refs.~\cite{gacmajid,kowaMEASURE},
while showing robustly that integration over energy-momentum is affected,
are insufficient to fully specify the new rules of integration.
Still the arguments reported in Refs.~\cite{gacmajid,kowaMEASURE}
imply that the net effect should be describable in terms of
a deformed measure of integration of the type
\begin{equation}
d^4p\rightarrow e^{\theta E/E_p}d^4p,\label{misuradeformata}
\end{equation}
with $\theta$ a parameter that will need to be adjusted on the basis of  future
better insight on the mechanism, but should most likely be
an integer multiple of $\eta$, with one of the values $\eta,-\eta,-3\eta$,
according to Ref.~\cite{gacmajid},  or the value  $3 \eta$,
according to Ref.~\cite{kowaMEASURE}.

In our Planck-scale modification of the Chandrasekhar model
we shall take into account both of modifications of the dispersion relation and
of modifications of the integration measure.

In closing this section let us give a simple illustrative derivation
of deformed integration measure that follows from a modified law of
composition of momenta.
For this illustrative example we focus
on the following
modified law of
composition of momenta
\begin{equation}
p \dot{+} q=\{E(\vec{p})+E(\vec{q}),\vec{p}+e^{-\eta E(\vec{p})/E_p}\vec{q}\} ~,
\label{psum}
\end{equation}
which is in use in the
literature on the $\kappa$-Minkowski noncommutative
spacetime~\cite{kappaM1,kappaM2,gacmajid,kappaM3}.
We look for the implications of this modified law of composition of momenta
for the function $F'(p)$ which is the integrand of $F(p)$: $F=\int F' dp$. And let us observe
that the law of composition of momenta (\ref{psum}) suggests that, for each spatial component
of momentum,
\begin{eqnarray}
F'(p)=\frac{F(p\dot{+}\Delta p)-F(p)}{\Delta p}=\frac{F(p+\Delta p e^{-\eta E/E_p})
-F(p)}{\Delta p}\simeq
 \frac{\partial F(p)}{\partial p}e^{-\eta E/E_p} ~.\label{rapinc}
\end{eqnarray}

This in turn suggests that in one spatial dimension one should have
\begin{eqnarray}
F=\int F' dp = \int F' dp = \int \frac{\partial F(p)}{\partial p}e^{-\eta E/E_p} dp~,\label{intONEDIM}
\end{eqnarray}
amounting effectively to a change of integration measure with
respect to the standard case: $dp \rightarrow e^{-\eta E/E_p}dp$. In the case here of interest,
with 3 spatial and 1 time dimension, one ends up with
\begin{equation}
d^4p \rightarrow e^{-3\eta E/E_p}d^4p ~,
\end{equation}
which is indeed compatible with (\ref{misuradeformata})
for $\theta = -3 \eta$.

\subsection{Combining modified dispersion relation and deformed momentum composition}
In light of the observations reported in the previous section we are interested
in an analysis of somewhat broader scope than the ones reported in Refs.~\cite{Camacho,sethNANE},
contemplating the implications for the
Chandrasekhar model of both a modification of type (\ref{MDR})
of the dispersion relation and of a modification of the law of composition of momenta
such that (on the basis of the observations reported in the preceding section,
and approximating $e^{\theta E/E_P}\simeq (1+\theta {E}/{E_p})$)
\begin{equation}
\left(1+\theta\frac{E}{E_p}\right)d^4p ~.\label{misuradeforme}
\end{equation}
This deformation of the integration measure, which results from a corresponding deformation
of the law of composition of momenta, affects the Chandrasekhar description of white dwarfs already
at the level of the relationship between the Fermi momentum $p_F$,
the number  $N$ of fermions in the system and  the volume $V$ of the star:
\begin{equation}
N=\frac{2V}{(2\pi)^3}4\pi\int_0^{p_F}p^2 \left(1+\theta\frac{E}{E_p}\right)dp ~.
\label{defpieffe}
\end{equation}
In particular, in the ultrarelativistic limit this leads to
\begin{equation}
p_F=\left(\frac{3}{4}\frac{N}{V}(2\pi)^2\right)^{\frac{1}{3}}\left(1
-\frac{1}{4}\frac{\theta}{E_p}\left(\frac{3}{4}\frac{N}{V}(2\pi)^2\right)^{\frac{1}{3}}
\right)\label{pfultrarel}~.
\end{equation}
Here the Planck-scale correction is such that for positive $\theta$
it decreases (with respect to the case without Planck-scale corrections)
the value of the Fermi momentum at a given density, or equivalently one can describe the
effect as going in the direction of allowing,
for given value of the Fermi momentum, to
place more particles in a given volume.
The opposite holds for negative $\theta$, which is a case with
increased value of the Fermi momentum at a given density, or lower density
at given Fermi momentum.

For the determination of the zero-point energy both the modification of the dispersion relation
and the modification of the integration measure play a role:
\begin{eqnarray}
E_0&=&\frac{2V}{h^3}\int_0^{p_F}  dp\left(1+\theta\frac{\sqrt{p^2
+m_e^2}}{E_p}\right) 4\pi p^2\sqrt{p^2+m_e^2} \left(1+\frac{\eta}{2}\frac{p^2}{p^2+m_e^2}\left(\frac{\sqrt{p^2
+m_e^2}}{E_p} \right)\right) ~. \label{Ezerocorr1nostra}
\end{eqnarray}

The Planck-scale features we are introducing in the Chandrasekhar model are all encoded
in the Fermi momentum and in the zero-point energy.
Let us focus again here on the ultrarelativistic behavior.
Concerning the equations of state one
straightforwardly finds\footnote{Some of the results in this section
are given in terms of  the mass $m_f$
of some fermions.
Within the context of the Chandrasekhar model, on which we focus in this section, $m_f$
should be taken as the electron mass, but in Section VI we shall refer to these formulas again,
and within the perspective of Section VI $m_f$ could still be the mass of electrons, but could also
be the mass of neutrons.}:
\begin{equation}
\epsilon(x_F)=\frac{E_0}{V} = \frac{3}{2}K\left(x_F(1+2x_F^2)\sqrt{1+x_F^2}
-\ln(x_F+\sqrt{1+x_F^2})+\frac{1}{5}\frac{m_f}{E_p}((\theta
+\frac{\eta}{2})x_F^5+\theta x_F^3)\right)\label{equazstatoA} ~,
\end{equation}
\begin{equation}
 P_0(x_F)=- \frac{\partial E_0}{\partial V} = \frac{1}{2}K\left( x_F(2x_F^2-3)\sqrt{1+x_F^2}
+\ln(x_F+\sqrt{1+x_F^2})+\frac{16}{5}\frac{m_f}{E_p}(\theta
+\frac{\eta}{2})x_F^5\right) ~,\label{equazstatoB}
\end{equation}
where $K=\frac{m_f^4}{12\pi^2}$, $\epsilon$
is the energy density, $P_0$ is the pressure of the fermionic gas, and,
also taking into account (\ref{pfultrarel}),
\begin{eqnarray}
x_F&\equiv&\frac{p_F}{m_e}=\frac{\bar{M}^{\frac{1}{3}}}{\bar{R}}\left(1
-\frac{\theta}{4}\frac{m_e}{E_p}\frac{\bar{M}^{\frac{1}{3}}}{\bar{R}}\right)
~.\label{xf}
\end{eqnarray}

These observations allow us to derive
a form of the pressure-balance equation
which takes into account
 both the modification of the dispersion relation and the modification
of  the law of composition of momenta
\begin{eqnarray}
K\frac{\bar{M}^{4/3}}{\bar{R}^4}\left(1
-\theta\frac{m_e}{E_p}\frac{\bar{M}^{\frac{1}{3}}}{\bar{R}}\right)
-K\frac{\bar{M}^{2/3}}{\bar{R}^2}\left(1
-\frac{\theta}{2}\frac{m_e}{E_p}\frac{\bar{M}^{\frac{1}{3}}}{\bar{R}}\right)
+\frac{8}{5}(\theta+\frac{\eta}{2})\frac{m_e}{E_p}K\frac{\bar{M}^{5/3}}{\bar{R}^5}
=K'\frac{\bar{M}^2}{\bar{R}^4} ~, &\label{equazdeform}
\end{eqnarray}
where on the left-hand side we used Eq.~(\ref{equazstatoB}),
while the right-hand side describes the pressure due to the Newtonian potential,
as done in Eq.~(\ref{Pgrav}).

From Eq.~(\ref{equazdeform})
one straightforwardly obtains (also see the analogous derivation in Section~II)
a relationship between radius and mass of the white dwarf, which is most
insightfully described in terms of a correction to the analogous relationship
that holds in the standard Chandrasekhar model:
\begin{eqnarray}
R=R_{Chan}\left(1+\left(\frac{13~\theta
+ 4~\eta}{10} \right)\frac{m_e}{E_p}\frac{1}{\left(1
-(\frac{M}{M_0})^{\frac{2}{3}}\right)^{\frac{3}{2}}}\right).\label{Camachos2}
\end{eqnarray}
This result, which characterizes our variant of the Chandrasekhar model
in the ultrarelativistic regime,
exactly reproduces Camacho's formula (\ref{Camachos})
upon taking $(13/4~\theta + \eta) \rightarrow \eta$.
The introduction of a deformed measure of integration, required in frameworks
with a modified law of composition of momenta, has not produced stronger
or different effects, but it is noteworthy that the effects due to the deformed measure
of integration, with parameter $\theta$, are exactly of the same order of magnitude
of the ones induced by the modification of the dispersion relation.
This implies that in particular frameworks with only a modified law of composition of
momenta (but no modification of the dispersion relation) would still predict
exactly the same features. And on the other hand in frameworks with both
a modified dispersion relation and a modified law of composition of momenta
one cannot a priori exclude a cancellation of the main Planck-scale effect
(the cancellation would occur if $\theta = - 4/13~\eta$).

In closing this section we observe that
by casting the result in the form (\ref{Camachos2})
(which we here adopted for allowing quick comparison
to the results of Ref.~\cite{Camacho})
some of the implications might be overlooked.
The form of Eq.~(\ref{Camachos2}) underlines that the effects
become significant when the mass of the white dwarf gets very close to
the Chandrasekhar mass, and this is a characterization that
provides little intuition for the type of features that should be sought in order
to find an amplification of Planck-scale effects in contexts that
are different from the one of the study of white dwarfs.
We notice however that Eq.~(\ref{Camachos2}) can be rewritten
equivalently as follows
\begin{eqnarray}
R=R_{Chan}\left(1+\frac{3}{2}\left(\frac{13\theta+4\eta}{10}\right)\frac{\pi^2}{m_p
m_e^2 E_p}\rho_{Chan}\right) \, ,   \label{Camachos3}
\end{eqnarray}
where $\rho_{Chan}=M/(4\pi R_{Chan}^3/3)$, and this provides the intuition that the effects are amplified
in presence of ultra-high densities ($ \rho_{Chan}\sim 2m_p m_e^2 E_p/(3\pi^2)$).
There is a certain ``quantum-gravity folklore" assuming that strong effects
could only arise for systems of Planck-length size, but this result
(in spite of its ``mere academic" nature) provides encouragement for a new intuition
according to which one could look for systems of extremely high density
(perhaps very-unusually high density) but not necessarily Planck-length size.

\section{Chandrasekhar model with IR modifications of the dispersion relation}\label{sectionIR}
Our generalization of the results of Refs.~\cite{Camacho,sethNANE}
exposed some possible differences between scenarios with broken Lorentz
invariance and scenarios with deformed Lorentz invariance and also 
provided some insight on a possible high-density regime of quantum gravity,
but still focused on the ultrarelativistic regime, where, as stressed above,
the phenomenology
of white dwarfs does not appear to offer any opportunities.
The
effects of the Planck-scale corrections become significant only for very small
white-dwarf radius, and for real stars of such small radius and
high densities the Chandrasekhar model is completely inapplicable.
The main opportunities
for a phenomenology based on the Chandrasekhar model (or its generalizations)
evidently are in a regime where particles are not ultrarelativistic, and that is
an awkward match for the standard ultraviolet/Planck-scale effects that are
most popular in the quantum-gravity literature, such as the ultraviolet modifications
of the dispersion relation and of the law of composition of momenta
we considered so far.

We notice however that there is a part of the quantum-spacetime literature where
analogous modifications of the dispersion relation have been studied, with implications
also when particles are not ultrarelativistic. This is the class of models
with ``ultraviolet/infrared (UV/IR) mixing".
There are several arguments suggesting that quantum gravity should
also have implications in some far-infrared regimes,
as was perhaps most eloquently advocated in Ref.~\cite{cohenUVIR},
on the basis of
our present understanding
of (quantum-) black-hole thermodynamics.
And this finds some
support in parts of the quantum-gravity/quantum-spacetime literature.
Of particular interest for our purposes is the case of modifications
of the dispersion relation which in the long-wavelength regime take the form
\begin{equation}
E^2 \simeq m^2 + p^2 - \xi p~.
\label{mdrurrutia}
\end{equation}
These were found in a quantum-spacetime model inspired by one of the competing perspectives
on the  (still unknown)
semiclassical limit of Loop Quantum Gravity~\cite{urrutiaPRLePRD} and is also the
qualitative behaviour\footnote{Light-like
noncommutativity with UV supersymmetry produces~\cite{szaboREVIEW}
corrections to the dispersion relation with IR behaviour
of type $\log (1+p_*/m)\propto p_*$ (where $p_*$ is the
spatial momentum in a preferential direction determined by the noncommutativity matrix).}
of the dispersion relation in the IR regime
found~\cite{szaboREVIEW} in the study of spacetime noncommutativity in the so-called ``light-like
noncommutativity models" (assuming UV supersymmetry~\cite{szaboREVIEW}).
Of course, one does not expect the energy scale $\xi$, introduced to characterize
the long-walvelength regime,
 to be ``Planckian", and indeed the relevant models encourage the expectation that $\xi \ll 1/L_P$.
The scale of onset of IR effects for light-like noncommutativity can be
described~\cite{szaboREVIEW}
 as a ratio
formed with an energy scale $M_{NC}$ that characterizes the noncommutativity matrix $\theta_{\mu \nu}$
($\theta_{\mu \nu} = - i [x_\mu ,  x_\nu]$)
and a much higher cutoff energy scale $\Lambda$, suggesting $\xi \sim M^2_{NC}/\Lambda$.
A somewhat similar mechanism favors small values of the energy scale $\xi$
also within the relevant Loop-Quantum-Gravity-inspired
model~\cite{urrutiaPRLePRD,josePRD,gacFlavioPRL2009}.

The understanding of UV/IR mixing is still at a rather preliminary
stage, and
Eq.~(\ref{mdrurrutia}) is only one among several candidate features that are being
considered. But, consistently with the exploratory nature of our study, we shall be satisfied
using Eq.~(\ref{mdrurrutia}) as an illustrative example within which we can articulate
some ideas on the possible relevance of UV/IR mixing for the study of white dwarfs
and other stars.

We start by analyzing the ``nonrelativistic" limit of
the Chandrasekhar model, still adopting the UV-modified dispersion relation.
Then we show how the analysis changes upon adopting (\ref{mdrurrutia})
as a candidate UV/IR-modification of the dispersion relation.

\subsection{Nonrelativistic limit of the modified Chandrasekhar model}
So we start with the task of further generalizing the analysis proposed by
Camacho by considering also the nonrelativistic regime.
For this we must first establish
the relationship between $x_F$, $\bar{M}$
and $\bar{R}$ in the nonrelativistic  limit.
We find appropriate to first do this
adopting the UV-modified dispersion relation, postponing to the next subsection
the study of the implications of (\ref{mdrurrutia})
as a candidate UV/IR-modification of the dispersion relation.

From Eq.~(\ref{defpieffe}), assuming indeed $x_F \ll 1$,
one finds
\begin{equation}
p_F=\left(\frac{3}{4}\frac{N}{V}(2\pi)^2\right)^{\frac{1}{3}}\left(1
-\frac{1}{5}\frac{\theta}{E_p}\frac{(\frac{3}{4}\frac{N}{V}(2\pi)^2)^{\frac{2}{3}}}{2
m}\right),\label{pfnonrel}
\end{equation}
which leads to
\begin{equation}
x_F=\frac{\bar{M}^{\frac{1}{3}}}{\bar{R}}\left(1
-\frac{1}{10}\frac{m_e}{E_p}\frac{\bar{M}^{\frac{2}{3}}}{\bar{R}^2}\right)~.
\end{equation}
This must be taken into account in the analysis of the
 zero-point energy, which
 is given by
\begin{equation}
P_0 =-\frac{\partial E_0}{\partial V} =\left(\frac{m_e^4}{15\pi^2}\right)\left(x_F^5
+(\eta+\theta)\frac{m_e}{E_p}x_F^5\right)~,\label{pizerononrelcorr}
\end{equation}
as easily found assuming $x_F \ll 1$ in Eq.~(\ref{Ezerocorr}).

We therefore find, combining this result (\ref{pizerononrelcorr})
with Eq.~(\ref{Pgrav}), that the Planck-scale-modified
 pressure-balance equation in the nonrelativistic limit takes the form
\begin{eqnarray}
\frac{4}{5}K\frac{\bar{M}^{\frac{5}{3}}}{\bar{R}^5}\left(1
+(\theta+\eta)\frac{m_e}{E_p}\right)-\frac{2}{5}\theta\left(\frac{m_e}{E_p}\right)K\frac{\bar{M}^{\frac{7}{3}}}{\bar{R}^7
}=K'\frac{\bar{M}^2}{\bar{R}^4},
\end{eqnarray}
which can be equivalently described as the following relationship
between mass and radius of the star:
\begin{eqnarray}
\bar{R}=\frac{4}{5}\frac{K}{K'}\bar{M}^{-\frac{1}{3}}\left(1+(\theta+\eta)\frac{m_e}{E_p}\right)-\frac{5}{8}\theta\left(\frac{m_e}{E_p}\right)\frac{K'}{K}\bar{M}~.
\end{eqnarray}
This result excludes the possibility of any meaningful phenomenological studies
of these effects in the nonrelativistic,  $x_F \ll 1$, regime,
since we are finding that the correction to the mass-radius formula is purely governed
by the ratio $m_e/E_p$ ($\sim 10^{-22}$).

\subsection{Illustrative example of Chandrasekhar model modified with UV/IR mixing}
We are now fully ready for exploring the implications
of the UV/IR-modified dispersion relation (\ref{mdrurrutia})
in the nonrelativistic regime of the Chandrasekhar model,
and we can quickly obtain a rather intriguing result.
We start by noticing that, assuming indeed $E^2\simeq m^2 + p^2-\xi p$,
and following the same procedure as in (\ref{Ezerocorr1}),
 one finds that the zero point energy of the system in this case is described by
\begin{eqnarray}
 E_0&=&\frac{2V}{h^3}\int_0^{p_F}4\pi p^2(\sqrt{p^2+m_e^2-\xi p}-m_e) dp ~.\label{Ezerocorr2}
\end{eqnarray}
From this it follows that
\begin{eqnarray}
E_0&=&\left(\frac{m_e^4 V}{\pi^2}\right)\left(f(x_F)-\frac{1}{6}\frac{\xi}{m_e}\left(2+(x_F^2-2)\sqrt{1+x_F^2}\right)\right).
\end{eqnarray}
In the non-ultrarelativistic regime ($x_F\ll 1$) this leads to the
following estimate of the IR-modified Pauli pressure
\begin{eqnarray}
 P_0=-\frac{\partial E_0}{\partial V}\simeq\left(\frac{m_e^4}{\pi^2}\right)\left(\frac{1}{15}x_F^5-\frac{1}{24}\frac{\xi}{m_e}x_F^4\right).
\end{eqnarray}
Equating Fermi pressure with Pauli pressure as in (\ref{equaz})
\begin{equation}
K'\frac{\bar{M}^2}{\bar{R}^4}=\frac{4}{5}K\frac{\bar{M}^{\frac{5}{3}}}{\bar{R}^5}-\frac{1}{2}K\frac{\xi}{m_e}\frac{\bar{M}^{\frac{4}{3}}}{\bar{R}^4},
\end{equation}
we can then obtain an IR-modified relationship
between mass and radius of the star:
\begin{equation}
\bar{R}\simeq\frac{4}{5}\frac{K}{K'}\bar{M}^{-\frac{1}{3}}
 \left(1-\frac{1}{2}\frac{K}{K'}\frac{\xi}{m_e}\bar{M}^{-\frac{2}{3}}\right).\label{mrIR}
\end{equation}
We should stress that here we have found no ``amplification" of the quantum-spacetime effects.
On the other hand, as stressed above, there are arguments suggesting that the
effects of UV/IR mixing should be generically ``small"  but without
a favored estimate of the smallness of the effects, so the ``amplification"
may not be necessary. And it is noteworthy that the qualitative behaviour of this
result is rather striking, as shown in Figure~1: for small (although not very small)
values of $\xi$ this UV/IR-modified Chandrasekhar model differs significantly from the unmodified one
in a region of sizes of the white dwarfs where data are relatively abundant.
And as shown in figure the nature of these modification is particularly interesting 
for the candidate ``strange white dwarfs"~\cite{NaneStrane}.\\
We shall not dwell here on how realistic the predictions of the specific setup we
considered as example of IR modification may be. That setup was only here considered
as an illustrative example, and we are here satisfied showing that
the qualitative implications are potentially interesting.

\begin{figure}[htbp!]
\includegraphics[width=\columnwidth]{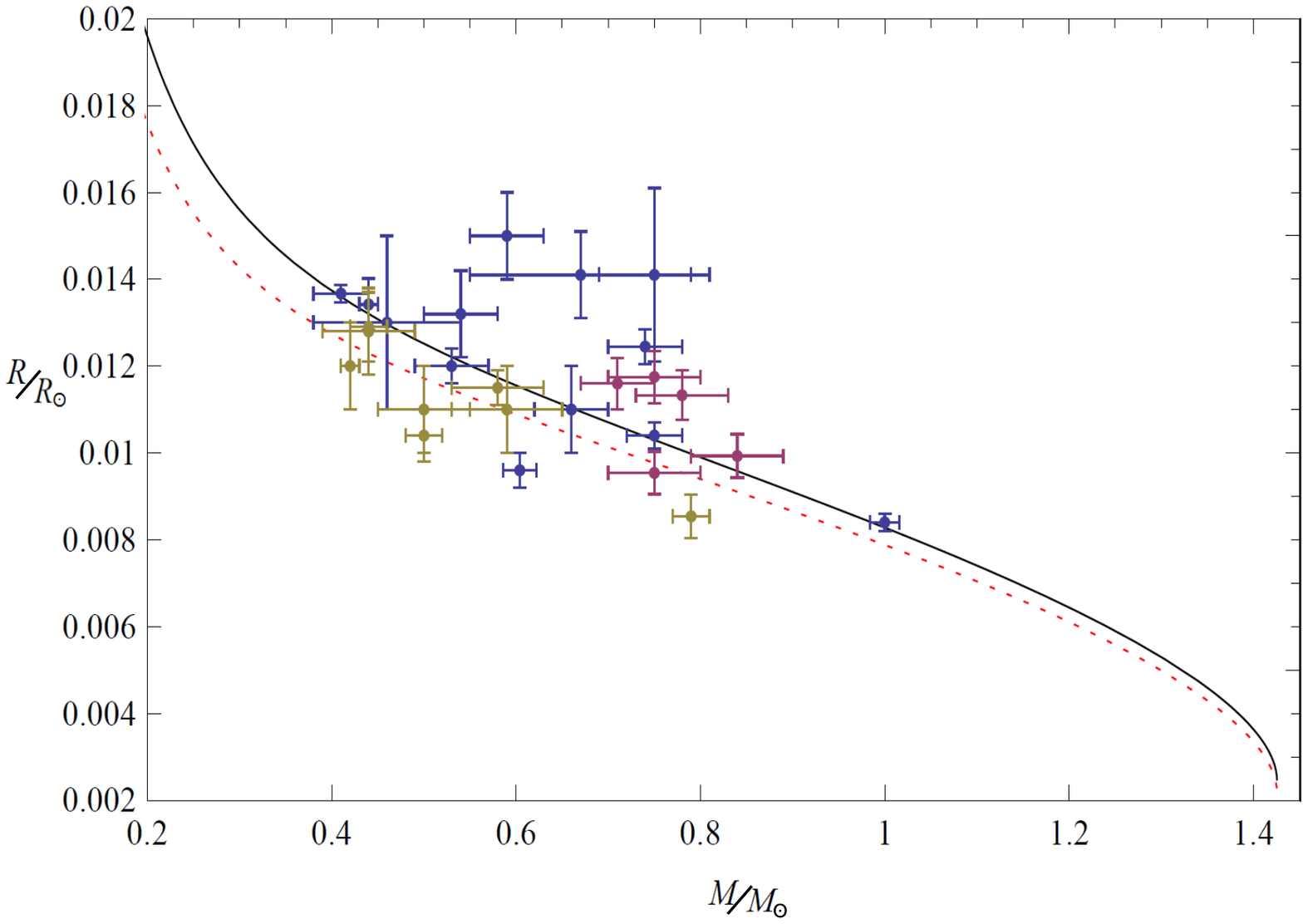}
\caption{In figure the black-continuous line describes the Chandrasekhar relationship
between mass and radius of white dwarfs (for helium).
Various factors, including composition (if different from Helium) and temperature, account for the fact that white dwarfs do not exactly fit the
Chandrasekhar line. The red-dashed line shows
our IR-modified result for the case of $\xi=0.15$. We show, as points with error bars,
some determinations of mass (in units of solar masses) versus radius (in units of solar radius)
for white dwarfs:  red points are the observed Praesepe white dwarfs~\cite{praesepe} and both the blue and the yellow points are from the Hipparcos survey~\cite{hypparcos1,hypparcos2}. In particular the yellow points have been identified as
 strange dwarfs candidates, partly as a result of the fact that composition of strange matter
 contributes to shifting the relationship
between mass and radius below the Chandrasekhar line for helium. The figure shows how one might fit
the yellow points on a IR-modified Chandrasekhar relationship of the type we here illustrated,
without invoking strange matter.}
\end{figure}

\section{High densities and deformed Tolman-Oppenheimer-Volkoff equations}\label{tolman}
Before closing let us return to the scenarios with ``standard" ultraviolet/Planck-scale
modifications of the dispersion relation. For the data available on white dwarfs
we have concluded that these cannot be relevant, and instead one could consider
scenarios with UV/IR-modified dispersion relations.
But our analysis with
ultraviolet/Planck-scale
modifications of the dispersion relation did provide encouragement for the idea
that in other contexts of interest these effects might become relevant,
if ultra-high densities are present.

In this brief section we provide further indication of this possible
role of high densities, and we also include some general-relativistic effects.
This we do within
the Tolman-Oppenheimer-Volkoff (TOV) equations~\cite{TOV1,TOV2}, which
provide a way for introducing general-relativistic
effects in the Chandrasekhar picture.

The starting point for our considerations
are the TOV equations\footnote{In this section we also set $G=1$}:
\begin{eqnarray}
\begin{cases}
\frac{dm}{dr}=4\pi\epsilon(P)r^2\\
\frac{dP}{dr}=-\frac{P+\epsilon(P)}{r(r-2m)}(4\pi Pr^3+m(r))
~.\end{cases}
\end{eqnarray}
Here $r$ is the distance from the center of the star, and the equations
simply give the general-relativistic description of the radial pressure gradient,
under the assumption of spherically symmetric mass distribution (so that $M = \int_0^R m(r) dr=\int_0^R
4 \pi r^2 \epsilon(r) dr$).

These TOV equations are solvable upon providing an equation of state, giving $\epsilon(P)$.
We shall of course be interested in the outcome of these TOV equations for the case
in which the equation of state is obtained from (\ref{equazstatoA})-(\ref{equazstatoB}),
so that it is an equation of state that takes into account our Planck-scale effects.

On the basis of the observations reported in the previous sections we expect
the ultraviolet/Planck-scale effects to be significant only in the ultrarelativistic regime,
and we therefore focus on solving\footnote{For $x_F \gg 1$ solving the TOV equations
is of little interest from a strict astrophysics perspective because of stability
concerns~\cite{Landau}.
From our perspective the analysis of the regime $x_F \gg 1$ of the TOV equations
is still meaningful, since we are here only interested in
estimating the possible magnitude of the effects.} the TOV equations for $x_F \gg 1$,
and for this purpose we find useful to make use of the variable
\begin{equation}
t=4\ln\left(x_F+\sqrt{1+x_F^2}\right)~.
\end{equation}
We also take into account that, as stressed in Section \ref{sectiontwo},
our Planck-scale effects of modification of the law of composition of momenta
introduce Planck-scale corrections, described in Eq.~(\ref{xf}),
for the formula for $x_F$ (whereas $x_F$ would not be affected by any Planck-scale
effects in scenarios that exclusively modify the dispersion relation).

In the ultrarelativistic regime ($x_F \gg 1$ and, as a result, also $t \gg 1$)
the TOV equations, if one indeed assumes that the Planck-scale-corrected $\epsilon(P)$
is obtained from (\ref{equazstatoA})-(\ref{equazstatoB}), take the form\footnote{For compactness,
in this section we work in units such that $K=2/(3\pi)$, which are not unusual in studies
of TOV equations. For a neutron gas
 this means that the unit of lenght has been fixed to
 be $l=1.36 \cdotp 10^4 m$, while the unit of mass
 is $\mu=1.83\cdotp 10^{31} Kg$.}
\begin{eqnarray}
\begin{cases}
\frac{dm}{dr}=\frac{1}{2}r^2e^t+\frac{1}{2}r^2(\frac{3}{2}\theta+\eta)\frac{m_f}{E_p}e^{\frac{5}{4}t}\\
\\
\frac{dt}{dr}=-\frac{4}{r(r-2m(r))}\left(\frac{r^3}{6}e^t+m(r)\right)
-\frac{r^3}{r(r-2m(r))}(\frac{23}{9}\theta+\frac{7}{4} \eta)\frac{m_f}{E_p}e^{\frac{5}{4}t}
\end{cases} \label{sistema3}
\end{eqnarray}

From these combined equations one can straightforwardly obtain
the explicit form of the $r$-dependence of $m$ and of (the exponential of) $t$
at leading order in $m_f/E_p$:
\begin{eqnarray}
e^{t(r)} &=&\frac{3}{7}\frac{1}{r^2}
+\frac{m_f}{E_p}\left(\frac{3}{7}\right)^{\frac{1}{4}}\left(\frac{884}{441}\theta
+\frac{271}{196}\eta\right)r^{-\frac{5}{2}} ~, \label{soluzione1}\\
m(r)&=&\frac{3}{14}r
+\frac{m_f}{E_p}\left(\frac{3}{7}\right)^{\frac{1}{4}}\left(\frac{2335}{882}\theta
+\frac{335}{196}\eta\right)r^{\frac{1}{2}} ~.\label{soluzione2}
\end{eqnarray}
These results can in turn be used
to establish the relation between mass and energy density, which is given by
\begin{equation}
m(\epsilon)\simeq \left(\frac{3}{14}\right)^{\frac{3}{2}}\epsilon_0^{-\frac{1}{2}}\left(1
+\frac{m_f}{E_p}f(\theta,\eta)\epsilon_0^{\frac{1}{4}}\right)~,\label{emmediro2}
\end{equation}
where $\epsilon_0$ is the central density of the star
 and  $f(\theta,\eta)=\frac{7}{3}\left(\frac{2777}{441}\theta+\frac{1611}{392}\eta\right)
+2^{\frac{1}{4}}(\frac{3}{2}\theta+\eta).$

It is noteworthy that our result (\ref{emmediro2}) shows that the Planck-scale effects
can be significant for ultra-high densities ($\epsilon_0 \sim (E_p/m_f)^4$),
without necessarily requiring the abstraction of a star of Planck-length size, an intriguing
possibility which was already encountered in our analysis of the modified Chandrasekhar model.

\section{Closing remarks}
The fact, established over the last decade, that there are some (however rare)
windows of opportunity for phenomenology of the quantum-spacetime/quantum-gravity realm represents a
(previously unexpected) significant
opportunity for this research area. Now that we have the first
few examples~\cite{gacPolonpapLRR}
of phenomenological
analyses establishing access to such effects, it is legitimate to get more ambitious
and  start worrying about the fact that
these first few examples only probe certain specific aspects of the quantum-gravity realm.
The study we here reported, taking off from related previous analyses reported in
Refs.~\cite{Camacho,sethNANE}, starts to set the stage for new ways to probe
the quantum-gravity realm in the context of macroscopic systems in astrophysics.

We unfortunately found that it is unlikely that near-term observations
of white dwarfs might play a role in the phenomenology of certain
ultraviolet/Planck-scale effects.
Yet the results of our analysis provides encouragement for the idea that
there is indeed a new regime of phenomena were tiny Planck-scale effects could
be observably large, and it is the regime of extremely high densities.
The densities that appear to be required for observably large Planck-scale effects
are ultra-high, and it might require a dedicated multi-stage research programme
to identify the most promising applications that could allow to
unveil such effects. Such a programme will need to face the challenges posed
by the still only partial ordinary-physics understanding of some of the
most interesting macroscopic systems in astrophysics, particularly when gravitational
collapse (and associated high densities) is involved. But the payoff that could be expected
appears to be well worth the effort, since such a novel window on the Planck-scale
realm could have particularly significant impact on our ability to investigate the
quantum-gravity problem.

We also believe that our observation concerning the possible relevance
of IR-modified dispersion relations (inspired by the mechanism of UV/IR mixing)
 for the understanding
of available data on white dwarfs should be of encouragement for related
further studies. The findings on that possibility that we reported
are clearly very preliminary and limited to a very specific model, used here only
for illustrative purposes. But
the qualitative comparison to data is encouraging enough to
warrant some dedicated studies, with more in depth analyses and considering a wider class
of scenarios for ``UV/IR mixing".

Another aspect of our analysis which should motivate further investigation
is the one concerning the possibility of
discrimination between scenarios with broken spacetime symmetries
and scenarios with deformed spacetime symmetries in the DSR sense.
As discussed in Section~\ref{sectiontwo}
through refinements of the analysis here reported
it should be possible to gain further insight on the type of observables that
can be used to discriminate between these two alternative scenarios for the fate of spacetime
symmetries at the Planck scale.

\end{document}